\documentstyle[twoside,fleqn,espcrc2]{article}

\newcommand\phibar{{\bar \phi}}

\newcommand{\AmS}{{\protect\the\textfont2
  A\kern-.1667em\lower.5ex\hbox{M}\kern-.125emS}}

\title{Fluctuation Induced First Order Phase Transitions}

\author{Yue Shen\address{Physics Department, Boston University,
                         Boston, MA 02115, USA}
        \thanks{This work was supported in part under DOE contract
DE-FG02-91ER40676 and NSF
contract PHY-9057173, and by funds from the Texas National Research Laboratory
Commission under grant RGFY92B6.}
       }
\begin{document}

\begin{abstract}
We study a $U(N)\times U(N)$ symmetric scalar field model in four and three
dimensions. First, using our data in four dimensions in the weak coupling
region, we demonstrate explicitly that
the observed first order phase transition is induced by quantum fluctuations.
Next, based on the renormalization group and our new simulation results in
three dimensions
we argue that even if the $U_A(1)$ symmetry is restored {\it below} the
critical temperature the QCD finite temperature chiral phase transition
for two flavor could be extremely weak first order.
\end{abstract}

\maketitle

\section{The Model}

Let us consider a scalar field model with action
\begin{eqnarray}
S_\phi &=&
 \int d^Dx \big\{ {1\over 2} tr \left(\partial_\mu \phi^\dagger\partial_\mu
\phi\right)
+ {m^2 \over 2}tr \left(\phi^\dagger\phi\right) \nonumber \\
&+&
\lambda_1\left(tr\phi^\dagger\phi\right)^2
+ \lambda_2 tr\left(\phi^\dagger\phi\right)^2
\big\} ~,
\label{eq:model}
\end{eqnarray}
where $\phi$ is a $N \times N$ complex matrix and $D$ is space-time dimension.
It has $U(N)\times U(N)$ global symmetry under transformation:
$\phi \to U\phi V^\dagger$, where $U$ and $V$ are $N\times N$ unitary matrices.

In contrast to a $O(N)$ scalar model, the symmetry breaking
phase transition (PT) for
Eq. (\ref{eq:model}) can be first order due to the
well known Coleman-Weinberg phenomenon \cite{Amit}.
A qualitative argument for the
Coleman-Weinberg phenomenon has been developed based on the effective
potential and the renormalization group (RG) \cite{Amit}.
Essentially first order PT occurs due to the existence of
run-away RG trajectories.
However, physical applications of Eq. (\ref{eq:model}) rely on the
RG flow obtained in perturbation \cite{BU} or $D=4 - \epsilon$ expansion
extrapolated to $\epsilon = 1$\cite{Pisarski}. It is important to extend
these arguments in a nonperturbative study.

\section{D=4}
Our first numerical results in 4D are reported in Ref. \cite{Shen},
which we refer to for details of the lattice simulations and
discussions of their relevance to the continuum physics.
As an illustration for the
quantum fluctuation induced first order PT, we plot in Fig. 1
a comparison of the one-loop perturbative calculation with MC data in the
weak coupling region. Here the one-loop effect changes the second order
PT at the tree level to first order.
First order PT were also found in the strong coupling
region \cite{Shen}, indicating the absence of a stable infrared fixed
point there.

\section{D=3}
In three dimensions, Eq. (\ref{eq:model})
can be used to describe the QCD finite temperature chiral phase
transition (ChPT) \cite{Pisarski}. Here the order of PT
depends crucially on the number of fermion flavor $N_f$ and the pattern
of symmetry. For $N_f \ge 3$ ChPT is in general
expected to be first order \cite{Pisarski}.
For $N_f=2$, if the axial $U_A(1)$ anomaly remains
effective up to the ChPT temperature $T_{ch}$, the symmetry of the
system will be $SU(2)\times SU(2) \sim O(4)$ and the transition can be
second order. However, $U_A(1)$ symmetry could be restored below
$T_{ch}$, the symmetry becomes $U_A(1)\times SU(2)
\times SU(2)$ and ChPT would be first order \cite{Pisarski}.

\begin{figure}[htb]
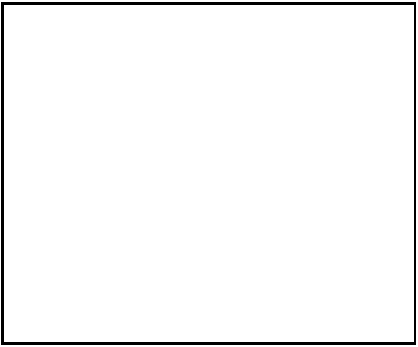

\framebox[55mm]{\rule[-21mm]{0mm}{43mm}}
\caption{Comparison of the numerical results with the bare perturbation
calculation (dotted line) in 4D. The numerical data are obtained on the $4^4$
(open circle), $6^4$ (open square) and $10^4$ (open triangle) lattices.
The solid lines connect $6^4$ and $10^4$ data points to indicate the
hysteresis effects.}
\label{fig:figure1}
\end{figure}

So far there is no numerical evidence for a first order ChPT
for $N_f=2$ \cite{Karsch}. Does this imply that
the $U_A(1)$ symmetry is not restored below $T_{ch}$? The answer can not be
conclusive. In fact, as we will show in the following using the effective
action in Eq. (\ref{eq:model}): even if we {\it assume} the $U_A(1)$ symmetry
is
restored below $T_{ch}$, ChPT can be very weakly first order.

For $N=2$ the RG flow for Eq. (\ref{eq:model}) is plotted in Fig. 2 where the
$\beta$-functions are obtained in $D=4-\epsilon$ expansion with $\epsilon=1$.
There is a UV fixed point at $\lambda_1=\lambda_2=0$ (solid circle)
and a unstable IR fixed
point at $\lambda_1 = \pi^2 \epsilon / 8,~\lambda_2 = 0$ (solid square).
The dotted lines in Fig. 2 indicate the ``stability line'' (SL). Qualitatively,
for a renormalized theory located to the left side of SL, PT
will be first order \cite{Amit}.
However, PT can be weakly or strongly first
order depending on the choice of the bare Lagrangian. The RG running of
$\lambda_1,\lambda_2$ is very slow in the weak coupling
region and becomes fast in the strong $\lambda_2$ region.
A weak coupling bare action chosen to the right side
of SL will need many decades of running in energy scale in order to cross SL.
Therefore, even though the PT will be first order, the jump
in the vacuum expectation value (VEV), $v$, at the transition point can be many
orders of magnitude smaller than
the ``cut-off'' scale $\Lambda$ and the transition can be very weakly first
order \cite{Shen}.

\begin{figure}[htb]
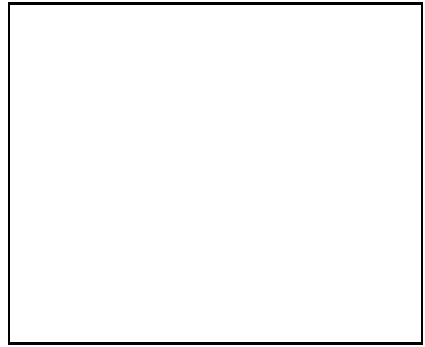

\framebox[55mm]{\rule[-21mm]{0mm}{43mm}}
\caption{RG flow in 3D. The arrows point to the infrared direction.}
\label{fig:figure2}
\end{figure}

As an example, we plot $2 v^2 = \langle tr\phibar^\dagger\phibar\rangle$ in
Fig. 3 for $\lambda_1 = \lambda_2=0.0625$. Obviously, it is almost impossible
to decide numerically at this point if the PT is second order
or weakly first order. Here we wish to correct an error appeared in
the literature \cite{Gaust}. The point at $m^2 = -6.8$ (solid circle in Fig. 3)
which was claimed to be
the critical point of a first order PT \cite{Gaust} is actually
deeply in the broken phase.

The search for a first order PT is further complicated
on a finite lattice. The running of $\lambda_1,\lambda_2$ is controlled
by the correlation length $\xi$, which is limited by the box size
$\xi \le L$. Therefore, a bare action fixed to the right of SL will only
be able to run across SL in a large enough box \cite{Shen}.
The first order nature
of the PT will not appear until the simulation is performed on
a large lattice. There is an interesting example of this phenomenon
in the literature \cite{Dreher}. In the simulation of $U(N)\times U(N)$
nonlinear sigma model, which is in the same universality class as
Eq. (\ref{eq:model}), it was found that for $N=2$ the first order nature
of the PT could be exposed only on lattices larger than $14^3$.
We should also point out that for the same initial values of $\lambda_1,
\lambda_2$, the RG running is faster with increasing $N$. This
explains the stronger first order PT for $N=3$ in Ref. \cite{Dreher}.

\begin{figure}[htb]
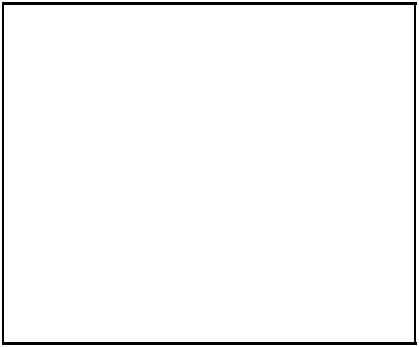

\framebox[55mm]{\rule[-21mm]{0mm}{43mm}}
\caption{In 3D, $\langle tr(\phibar^\dagger\phibar)\rangle $
as a function of $m^2$ at $\lambda_1 = \lambda_2 = 0.0625$.
The bare perturbation prediction is given by the solid line.}
\label{fig:figure3}
\end{figure}

\begin{figure}[htb]
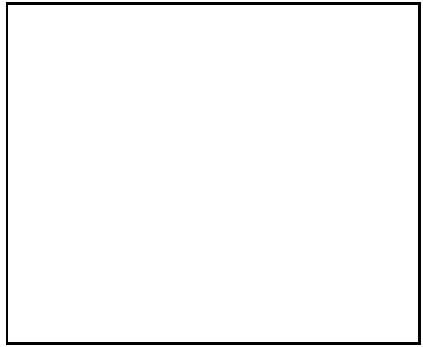

\framebox[55mm]{\rule[-21mm]{0mm}{43mm}}
\caption{In 3D, Monte Carlo time history for
$tr(\phibar^\dagger\phibar)$
at $\lambda_1 = -0.22, \lambda_2 = 0.5, m^2 = -1.038$.
Because of the closeness to the ``stability line'', we are able to
observe tunneling between two states on a $6^3$ lattice.}
\label{fig:figure4}
\end{figure}

To observe the first order PT in a small box, one either
has to move the bare action close to SL in the weak coupling region, or move
to the strong $\lambda_2$ region where the speed of running is fast
\cite{Shen}.
We give an example in Fig. 4 where the bare action is set to be
very close to SL.
We are able to observe clear signal for a first order PT on the
$6^3$ lattice.

What are the implications for finite temperature QCD?
Let us assume that Eq. (\ref{eq:model}) becomes a good effective theory
for QCD at energy scale $\Lambda < 2\pi \Lambda_{QCD}$.
Then the strength of first order PT
crucially depends on the values of the effective coupling
at scale $\Lambda$. As discussed above, if $\lambda_1(\Lambda),
\lambda_2(\Lambda)$ fall into the weak coupling region, finding the first
order PT will become a tough job. While it is not possible
to determine $\lambda_1(\Lambda), \lambda_2(\Lambda)$ for QCD, it was found in
a
large color ($N_c$) approximation for the Nambu-Jona-Lasinio (NJL) model that
$\lambda_1 = 0, \lambda_2 = 2\pi^2/N_c
\ln(\Lambda_{NJL}/\Lambda)$ \cite{Bardeen}
where $\Lambda_{NJL}$is the cut-off for NJL model and $\Lambda$ is the scale
when the linear sigma model becomes a good effective theory.
Although NJL model is a poor approximation of QCD, we might believe
that it is at least qualitatively correct.
Then $\lambda_2$ can be strong or
weak depending on the ratio $\Lambda_{NJL}/\Lambda$ (one might take
$\Lambda_{NJL} \sim 4\pi\Lambda_{QCD}$).
Since $\lambda_2$ does not depend on $N_f$, the only feature that distinguishes
flavor is the speed of RG flow.
We could immediately draw a conclusion
that it would be easier to observe the first order PT for
$N_f =4$ than $N_f=2$, which agrees with well known simulation
results \cite{Karsch}.
Even in the case of $4\pi\Lambda_{QCD}/\Lambda \sim 1$ and $\lambda_2$ is
very large (Eq. (\ref{eq:model}) becomes effectively
the nonlinear sigma model), results of
Ref. \cite{Dreher} indicate that one may need a quite large lattice to
expose the first order nature of PT.

Beyond large $N_c$ limit, one in principle can not eliminate the
possibility of a stronger $N_f$ dependence for QCD.
Then for $N_f=2$, $\lambda_2(\Lambda)$ could fall
in the very weak coupling region such that PT is extremely weak first order.

I thank R. Pisarski for useful discussions.

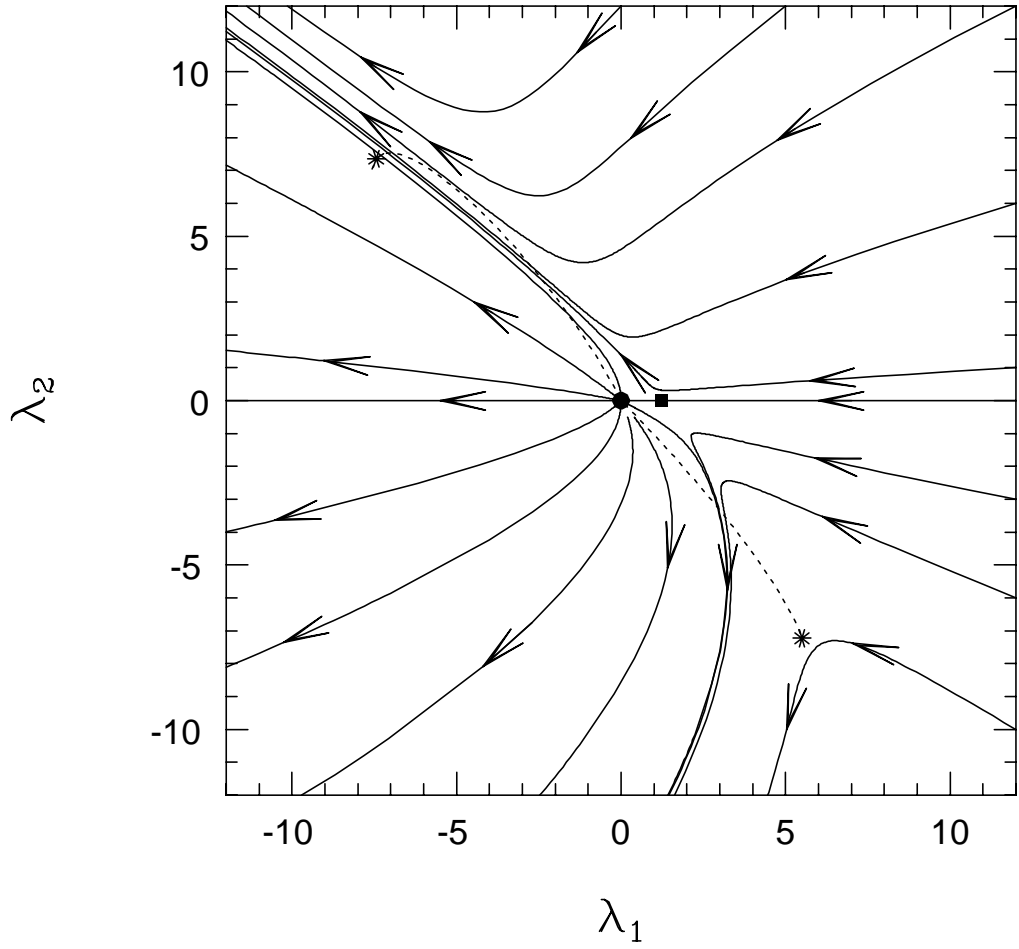
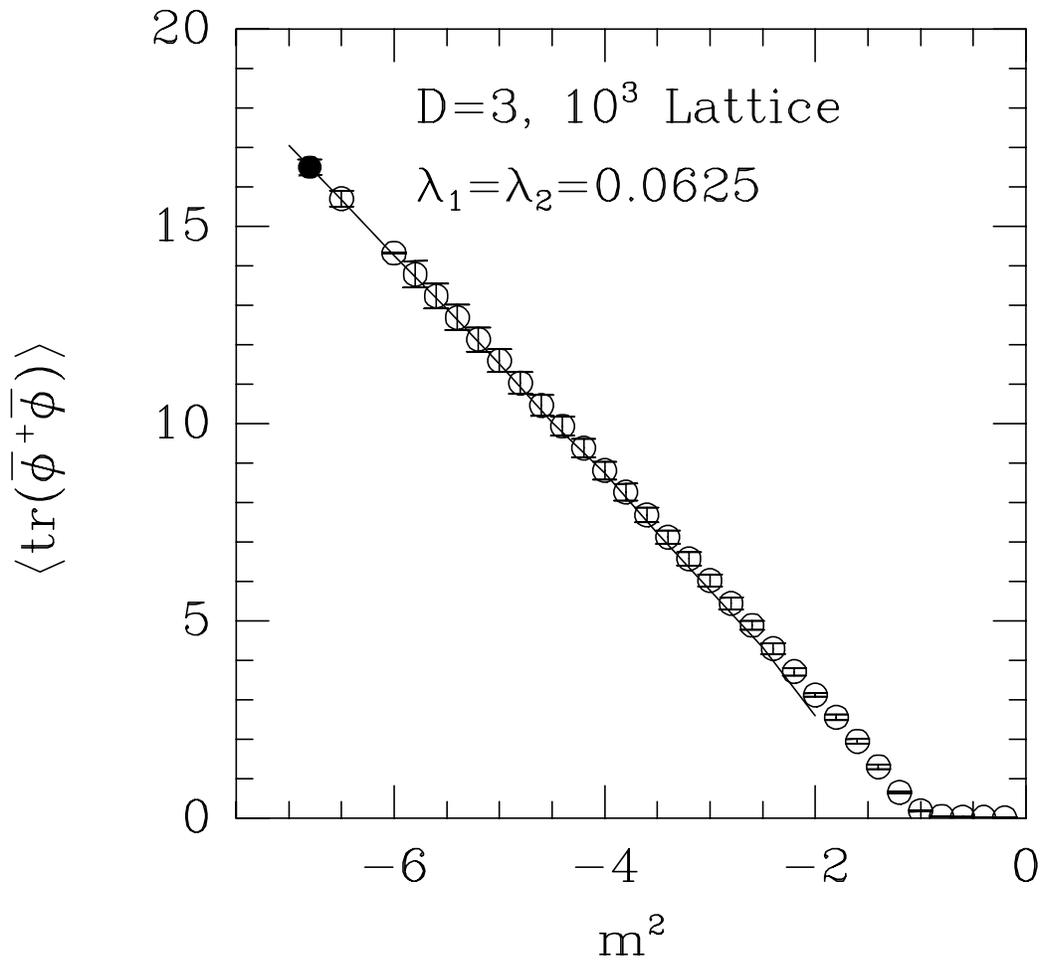
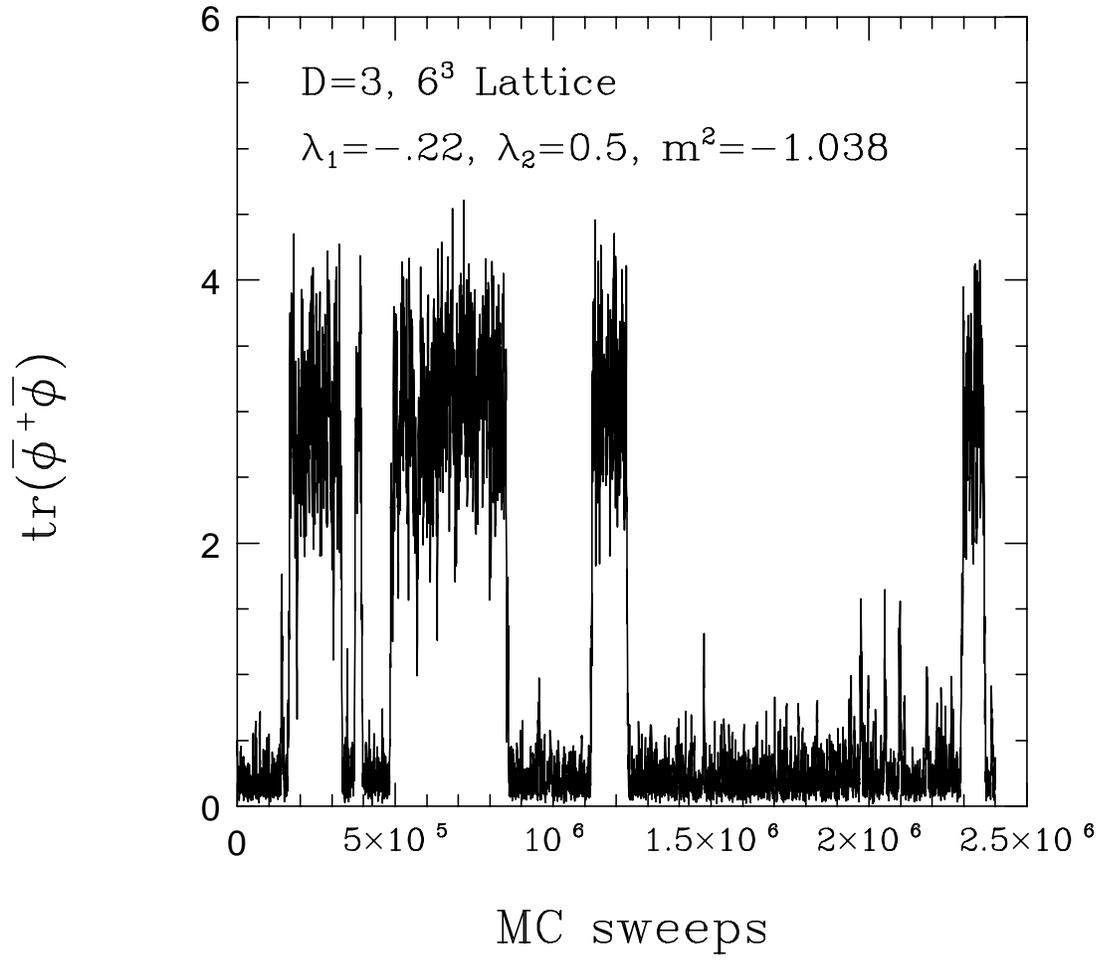
\end{document}